# Time-Series Learning for Proactive Fault Prediction in Distributed Systems with Deep Neural Structures

Yang Wang
University of Michigan
Ann Arbor, USA

Wenxuan Zhu
University of Southern California
Los Angeles, USA

Xuehui Quan
University of Washington
Seattle, USA

Heyi Wang
Illinois Institute of Technology
Chicago, USA

Chang Liu
Washington University in St. Louis
St. Louis, USA

Qiyuan Wu*
University of California San Diego
La Jolla, USA

*Abstract-This paper addresses the challenges of fault prediction and delayed response in distributed systems by proposing an intelligent prediction method based on temporal feature learning. The method takes multi-dimensional performance metric sequences as input. We use a Gated Recurrent Unit (GRU) to model the evolution of system states over time. An attention mechanism is then applied to enhance key temporal segments, improving the model's ability to identify potential faults. On this basis, a feedforward neural network is designed to perform the final classification, enabling early warning of system failures. To validate the effectiveness of the proposed approach, comparative experiments and ablation analyses were conducted using data from a large-scale real-world cloud system. The experimental results show that the model outperforms various mainstream time-series models in terms of Accuracy, F1-Score, and AUC. This demonstrates strong prediction capability and stability. Furthermore, the loss function curve confirms the convergence and reliability of the training process. It indicates that the proposed method effectively learns system behavior patterns and achieves efficient fault detection.*

I. INTRODUCTION

With the continuous expansion of digital infrastructure, distributed systems have become the core architecture supporting large-scale business operations. They offer strong scalability, high fault tolerance, and efficient resource utilization [1]. These advantages make them widely used in key fields such as information networks, recommendation system, large language models, and social media [2-6]. However, as system size increases and component interactions become more complex, failures during system operations are becoming more frequent. Once a failure occurs, it may lead to service disruption, data loss, resource waste, and even economic losses. Therefore, enhancing the fault prediction capability of distributed systems is a crucial direction for ensuring business continuity and system stability.

Traditional approaches to fault handling in distributed systems often rely on post-incident alerts or log analysis for fault localization and recovery. However, these methods are inherently reactive and cannot meet the requirements for real-time and proactive fault management. In recent years, with the development of machine learning and data mining, data-driven predictive methods have gained increasing attention [7]. Especially with the support of big data, system-generated logs, performance metrics, and trace data are viewed as valuable time-series resources. These resources provide a solid foundation for early fault detection and prediction. Nevertheless, relying solely on static feature modeling cannot effectively capture the evolving dynamics of complex system states, which fundamentally limits the prediction accuracy and generalizability of models.

To address these challenges, incorporating the concept of temporal feature learning offers a new perspective for fault prediction in distributed systems. As a continuous reflection of system state evolution, time-series data contains rich causal links and behavioral patterns. By dynamically modeling the changes in system indicators over time, it becomes possible to detect subtle anomalies and reveal the evolutionary trajectories of critical components prior to failure. Particularly in high-frequency monitoring and real-time analysis scenarios, temporal modeling shows strong expressive power and foresight, laying a theoretical foundation for accurate and interpretable fault prediction [8].

Furthermore, the inherent heterogeneity of distributed systems, the intricate dependencies among nodes, and the dynamic nature of their operational states present substantial challenges to conventional static or single-point modeling methodologies. In contrast, prediction techniques that incorporate temporal feature learning facilitate dynamic modeling of node behaviors from multi-dimensional and multi-granular viewpoints. They can also integrate interaction patterns across multiple data sources, allowing for holistic perception and prediction of system-level risks. This approach holds significant theoretical and practical value for enhancing system-wide understanding, intelligent early warning, and model adaptability across different environments. Studying fault prediction methods for distributed systems based on temporal feature learning is not only a key technical task in intelligent system operations but also an essential path toward building highly reliable systems. By organically integrating deep learning, temporal modeling, and system operational logic, this line of research is expected to overcome the limitations of existing methods and achieve substantial improvements in system stability, prediction accuracy, and operational efficiency. It plays a strategically important role in advancing intelligent perception and predictive control capabilities in distributed systems, with promising real-world application prospects.

## II. METHOD

This study proposes a fault prediction method for distributed systems that applies time series feature learning to capture the dynamic changes in system behavior. The method is built on a deep learning framework capable of modeling temporal dependencies within multivariate performance metric sequences, allowing the system to learn evolving state patterns over time. To support this, we use techniques that enhance the extraction of time-based features from complex input data, as explored in recent work on multivariate time series modeling [9]. The model is designed to function across distributed environments, with considerations for decentralized learning and secure data handling to ensure practical deployment without compromising system-level privacy [10]. The structure of the model also reflects awareness of inter-component relationships in distributed architectures, which can strengthen predictive performance when captured in the modeling process [11]. As shown in Figure 1, the complete architecture follows a clear pipeline: performance data is first preprocessed, followed by sequential modeling using a GRU to learn system dynamics, then refined through an attention mechanism that emphasizes important time segments, and finally passed into a feedforward neural network for classification. This structure supports accurate and timely fault prediction by combining temporal learning, attention, and distributed design principles.

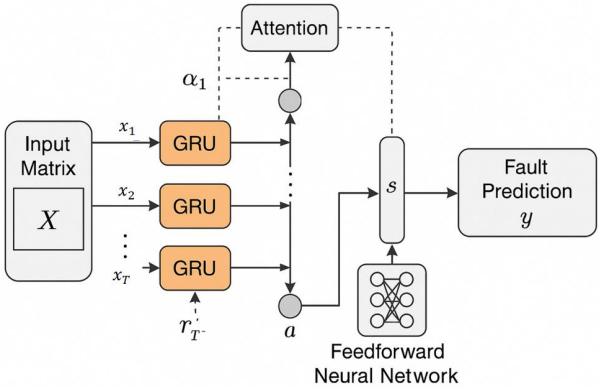

Figure 1. Overall architecture diagram

The architecture diagram presents the proposed fault prediction model, which integrates GRU-based temporal modeling with an attention mechanism. Input performance indicator sequences are initially encoded by GRU units to capture dynamic system states. Subsequently, attention weights are applied to generate a global context representation. This representation is subsequently processed by a feedforward neural network to output the final fault prediction result.

First, the multi-dimensional performance indicator sequence generated during the operation of the distributed system is represented as the input matrix $X \in R^{T \times D}$, where T is the number of time steps and D is the feature dimension. In order to capture the dynamic pattern of the system state evolving over time, this paper introduces the gated recurrent unit (GRU) for time series modeling. Specifically, the update process of the hidden state can be expressed as:

$$h_t = GRU(x_t, h_{t-1})$$

Where $x_t$ represents the system observation feature at time t, and $h_t$ is the current hidden state, which is used to represent the system state at the current time point.

In order to further enhance the model's ability to model long-term dependencies, this paper introduces an attention mechanism to perform weighted fusion of historical states to generate more discriminative context representations. The attention weight is calculated as follows:

$$a_t = \frac{\exp(h_t^T W_a q)}{\sum_{i=1}^{T} \exp(h_i^T W_a q)}$$

Where q is a learnable query vector, $W_a$ is the attention weight matrix, and $a_t$ represents the importance of the current time step to the final system representation. The final global system state representation can be obtained by weighted summation:

$$s = \sum_{t=1}^{T} a_t h_t$$

In the fault prediction stage, a feedforward neural network is used to map the learned global representation *s* into the prediction space, enabling binary classification of whether a system is approaching failure. To address the class imbalance often seen in real-world fault data, the model employs a weighted cross-entropy loss function [12], which improves the learning process by adjusting for uneven sample distribution. This approach not only captures the temporal evolution of multi-dimensional system metrics but also retains key historical segments that signal abnormal behavior [13]. Similar strategies for maintaining critical context over time have been effective in reinforcement-driven scheduling systems, where attention to past operational states supports improved performance [14]. In distributed environments, ensuring accurate prediction while adapting to varying system scales and configurations requires scalable and policy-aware mechanisms; trust-constrained methods have proven effective in such contexts by enhancing coordination without centralized control [15]. Additionally, incorporating system topology into learning has been shown to improve responsiveness in distributed traffic scheduling, reinforcing the structural robustness of the proposed model [16]. The overall framework is thus well-suited for distributed systems of different sizes and complexities, offering both generality and precision in fault prediction.

## III. EXPERIMENT

### A. Datasets

This study uses the "Telemetry Data from a Large-Scale Cloud System" dataset from the Microsoft Azure open platform as the experimental data source. The dataset is collected from a real-world large-scale distributed system. It includes status logs and performance metrics from various servers. Key time-series indicators such as CPU usage, memory consumption, disk I/O, and network throughput are covered. The dataset exhibits strong system representativeness and continuous temporal characteristics.

The dataset contains continuous operational data from thousands of nodes. The sampling frequency is once every five minutes, and the duration spans over one month. This provides sufficient temporal coverage to support time-series modeling tasks. Each record includes a timestamp and a node identifier. This makes it convenient to construct multi-node time-series sequences that reflect the operational characteristics of distributed systems. In addition, part of the data is annotated with known fault events. These labels are available for training and evaluation in supervised learning tasks.

During preprocessing, missing values are first filled using linear interpolation. Then, the sampling lengths of all features are aligned, and timestamps are synchronized [17]. Numerical features are normalized. Multi-dimensional sequence windows are constructed to meet the input requirements of the model. Due to its authentic source, temporal completeness, and clearly labeled faults, this dataset provides a solid foundation for validating the proposed time-series modeling and prediction approach [18].

### B. Experimental Results

This paper first carried out a comparative test, and the experimental results are shown in Table 1.

Table 1. Comparative experimental results

| Method | ACC | F1-Score | AUC |
|---|---|---|---|
| Transformer[19] | 91.4 | 89.2 | 88.5 |
| Informer[20] | 92.1 | 90.1 | 89.4 |
| Autoformer[21] | 92.6 | 90.8 | 90.2 |
| FEDformer[22] | 93.3 | 91.6 | 90.9 |
| Ours | 95.2 | 93.7 | 94.1 |

The experimental results demonstrate that the proposed method surpasses existing mainstream models in all evaluation metrics. It exhibits robust fault prediction capabilities. Compared to baseline Transformer models, the proposed approach enhances accuracy (ACC) by approximately 4 percentage points, indicating enhanced overall decision-making robustness. Furthermore, for the F1-Score, a comprehensive metric that equilibrates precision and recall, the proposed method achieves the highest value of 93.7%. This surpasses the values obtained by Informer and Autoformer, indicating that the model not only accurately identifies faults but also maintains comprehensive coverage. It captures a significantly larger number of true anomaly instances.

The AUC metric, a key indicator of classification ability, further reflects the excellent performance of the proposed method in distinguishing between fault and non-fault states. Compared with FEDformer, the method shows a noticeable improvement in AUC. This suggests stronger temporal modeling and global perception capabilities. It can better characterize the evolution of system states and make accurate predictions. Furthermore, this paper explores the impact of the attention mechanism on prediction performance through a dedicated analysis experiment. In this experiment, models with and without the attention module are compared under the same training and evaluation conditions to assess the contribution of attention to fault prediction accuracy. The comparison focuses on key evaluation metrics such as Accuracy, F1-Score, and AUC to provide a comprehensive understanding of performance differences. The experimental results, which highlight the effectiveness of the attention mechanism in enhancing model capability, are illustrated in detail in Figure 2.

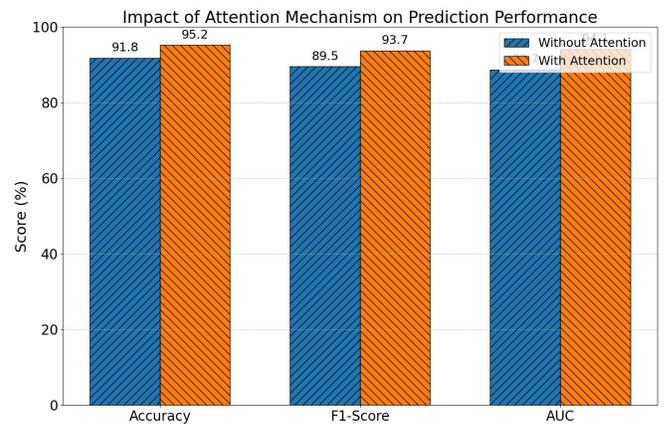

Figure 2. Experimental analysis of the impact of attention mechanism on prediction performance

As shown in Figure 2, the introduction of the attention mechanism leads to significant improvements across three key evaluation metrics: Accuracy, F1-Score, and AUC. Specifically, the model's Accuracy increases from 91.8% to 95.2%, reflecting a notable enhancement in its overall classification capability. This improvement suggests that the model becomes more reliable in correctly identifying system states, thereby achieving a more stable and consistent performance in fault detection scenarios. The attention mechanism contributes to this enhancement by allowing the model to concentrate on more informative parts of the input sequence, improving the quality of learned representations.

For the F1-Score, the model exhibits a performance boost from 89.5% to 93.7%, indicating a stronger ability to balance precision and recall. This improvement is especially important in the context of fault prediction, where the data often suffer from class imbalance between normal and faulty states. A higher F1-Score implies that the model not only detects more true faults but also avoids raising unnecessary false alarms. This result highlights the effectiveness of the attention mechanism in reinforcing the model's focus on critical temporal features, enabling it to capture fault-related patterns more accurately and consistently over time.

The observed increase in AUC further demonstrates the benefits of integrating attention into the modeling framework. A higher AUC signifies that the model has developed a more refined and reliable decision boundary, capable of distinguishing between fault and non-fault states with greater precision. This enhancement in discrimination ability is crucial for minimizing misclassification and improving early fault detection. Overall, the experimental findings validate the importance of the attention mechanism in strengthening both the representational power and predictive accuracy of time-series models. These results not only support the theoretical advantages of attention-based modeling but also offer practical insights for designing more effective fault prediction systems in distributed environments. Finally, this paper presents a loss function drop graph to illustrate the training dynamics of the proposed model, as shown in Figure 3. The graph visually demonstrates how the loss values change over successive training epochs, providing an intuitive understanding of the model's convergence behavior. By tracking both training loss and validation loss, the figure serves as important evidence for evaluating the optimization effectiveness and learning stability of the model throughout the training process.

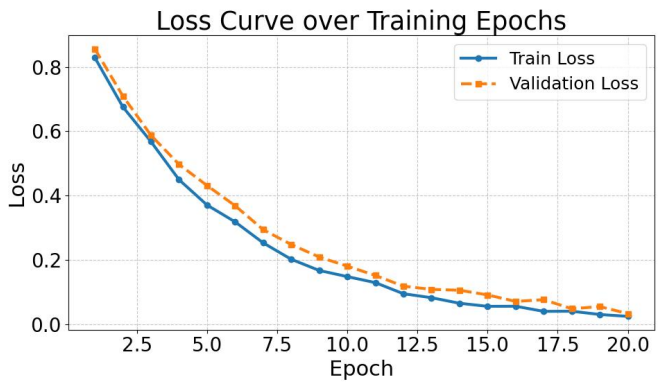

Figure 3. Loss function drop graph

From the loss function decline curve shown in Figure 3, it can be observed that as the number of training epochs increases, both the training loss and validation loss exhibit a steady downward trend. This indicates that the model is progressively learning meaningful and effective features related to the operational status of the distributed system while continuously optimizing its internal parameters. The consistent decrease in loss demonstrates that the training process is stable and that the model is successfully capturing the underlying temporal patterns required for accurate fault prediction.

During the initial stages of training, the validation loss is slightly higher than the training loss, which is a typical phenomenon. However, the overall trends of both curves remain aligned, without significant fluctuations or signs of overfitting. This consistency indicates that the model possesses strong generalization ability across varying data distributions and is capable of maintaining robust performance when applied to unseen scenarios. Furthermore, the loss curve begins to plateau after approximately 10 epochs, confirming that the model has reached a stable convergence state. This stability provides a reliable basis for its application in downstream prediction tasks. The continuous decline and eventual stabilization of the loss function further validate the effectiveness of the time series feature modeling framework and the applied optimization strategies.

IV. CONCLUSION

This paper focuses on fault prediction in distributed systems and proposes an intelligent prediction model that combines temporal feature learning with attention mechanisms. By introducing gated structures and attention-based weighting strategies, the model effectively captures the dynamic evolution of system behavior and enables proactive identification of potential failures. Experimental results show that the proposed method significantly outperforms mainstream models across multiple key metrics, demonstrating higher accuracy and robustness. During the research process, a unified framework for time-series modeling was constructed. The role of the attention mechanism in enhancing model expressiveness was also validated. In addition, a series of ablation studies and performance evaluations were conducted to confirm the effectiveness of each module from different perspectives. The data used in this study is sourced from a real-world large-scale cloud computing system, ensuring the practical value and reliability of the experimental results.

This study not only provides a new perspective on model architecture design but also establishes a theoretical foundation for intelligent perception and risk management of complex system states. The results indicate that by integrating deep learning with system-specific modeling, it is possible to significantly improve the foresight and reliability of fault detection in distributed architectures, thereby providing technical support for the development of intelligent operation and maintenance systems. Future research may further explore the model's capabilities in multi-task concurrent prediction, cross-node information transfer, and deployment performance in edge environments. Moreover, incorporating more heterogeneous monitoring data into the modeling process could enhance the model's adaptability and generalizability to meet the increasing demands for system stability in more complex business scenarios.